\documentclass[11pt]{article}

\usepackage[margin=1in]{geometry}
\usepackage{graphicx}
\usepackage{booktabs}
\usepackage{amsmath}
\usepackage{natbib}
\usepackage{hyperref}
\usepackage{xcolor}
\usepackage{float}
\usepackage{array}

\title{An In-Silico Audit of Topographic Organization in a Multimodal Brain-Response Model}

\author{%
  Jakub Strawa \\
  StrawAI, Krakow, Poland \\
  \texttt{jakubstrawaai@gmail.com}
}

\date{\today}

\begin{document}
\maketitle

\begin{abstract}
TRIBE v2 \citep{dascoli2026tribe} predicts fMRI responses across the cortical surface from
video, audio and text. Its authors validated it with category-selective localizers (faces,
places, bodies, words) and language contrasts, but not with the classic topographic
organization of sensory cortex: retinotopy, tonotopy, and motion selectivity in area MT+.
There is a clear architectural reason to expect retinotopy to fail: TRIBE's video features
are averaged over space before the response head, which should discard visual-field
position. I present mapping stimuli, a fitting pipeline adapted from standard fMRI mapping
analyses, and a stand-in model with known ground truth that serves as a positive and
negative control, and apply them to TRIBE v2's predictions, comparing the fitted maps with
human atlases using spin tests. TRIBE v2 shows a sharp split. There is no contralateral
hemifield response and no polar-angle map, robustly across stimulus controls. There is no
tonotopic gradient in primary auditory cortex: best frequencies span an interquartile range
of only 0.14 octaves across core and belt, and the null persists with loudness-equalized
tones. An eccentricity-like gradient appears only when the ring geometry changes, consistent
with global image statistics rather than a positional code. In contrast, motion selectivity
is strong and precisely localized: 22\% of the 1\% most motion-selective vertices fall in
MT+, which covers 0.42\% of cortex. In-silico experiments that depend on feature identity
appear viable with TRIBE v2, while those that depend on visual-field or frequency position
are not supported by the current model.
\end{abstract}

\section{Introduction}

TRIBE v2 \citep{dascoli2026tribe} is a foundation model trained to predict voxel/vertex-wise
fMRI responses on the fsaverage5 cortical surface (20{,}484 vertices) from naturalistic
video, audio and text input. Its original evaluation covered category-selective visual
localizers (faces $\to$ FFA, places $\to$ PPA, bodies $\to$ EBA, words $\to$ VWFA) and
language contrasts, where it performed well. It did not evaluate three of the
best-replicated organizational principles in sensory neuroscience: retinotopy in early
visual cortex, tonotopy in primary auditory cortex, and motion-selective responses in area
MT+.

There is a specific architectural reason to expect retinotopy to fail: the paper's own
methods describe averaging V-JEPA~2 patch tokens over space before they reach the response
head, which by construction removes the positional information a polar-angle or
eccentricity map requires. No prior work has quantified whether this prediction holds, or
whether motion selectivity (which does not depend on position, so should survive spatial
averaging) and tonotopy (whose audio features are not pooled in the same way) fare
differently.

I test four hypotheses, stated before running the model (Table~\ref{tab:hypotheses}):

\begin{table}[h]
\centering
\begin{tabular}{p{0.18\linewidth}p{0.38\linewidth}p{0.34\linewidth}}
\toprule
& \textbf{Prediction} & \textbf{Rationale} \\
\midrule
\textbf{H1} polar angle / contralaterality & Absent: no contralateral V1 response to
  hemifield stimuli, no polar-angle map & Spatial averaging of video features \\
\textbf{H2} eccentricity & Partly present: an eccentricity-like gradient from global image
  statistics & Averaged features still change with ring size \\
\textbf{H3} motion & Present: moving $>$ static dots peaks in MT+ & V-JEPA~2 is a video
  model; the original paper found a motion-related component \\
\textbf{H4} tonotopy & Open: best-frequency gradient in A1 & Audio features are not pooled
  over frequency, but training data were mostly speech/film sound \\
\bottomrule
\end{tabular}
\caption{Hypotheses, stated before running the model.}
\label{tab:hypotheses}
\end{table}

Any outcome is informative: confirming H1 quantifies a concrete limitation for anyone using
TRIBE for in-silico experiments that depend on visual-field position; disconfirming it would
be a surprising finding about what spatially-averaged transformer features retain.

\section{Methods}

\subsection{Stimuli}
\begin{figure}[H]
\centering
\includegraphics[width=0.85\linewidth]{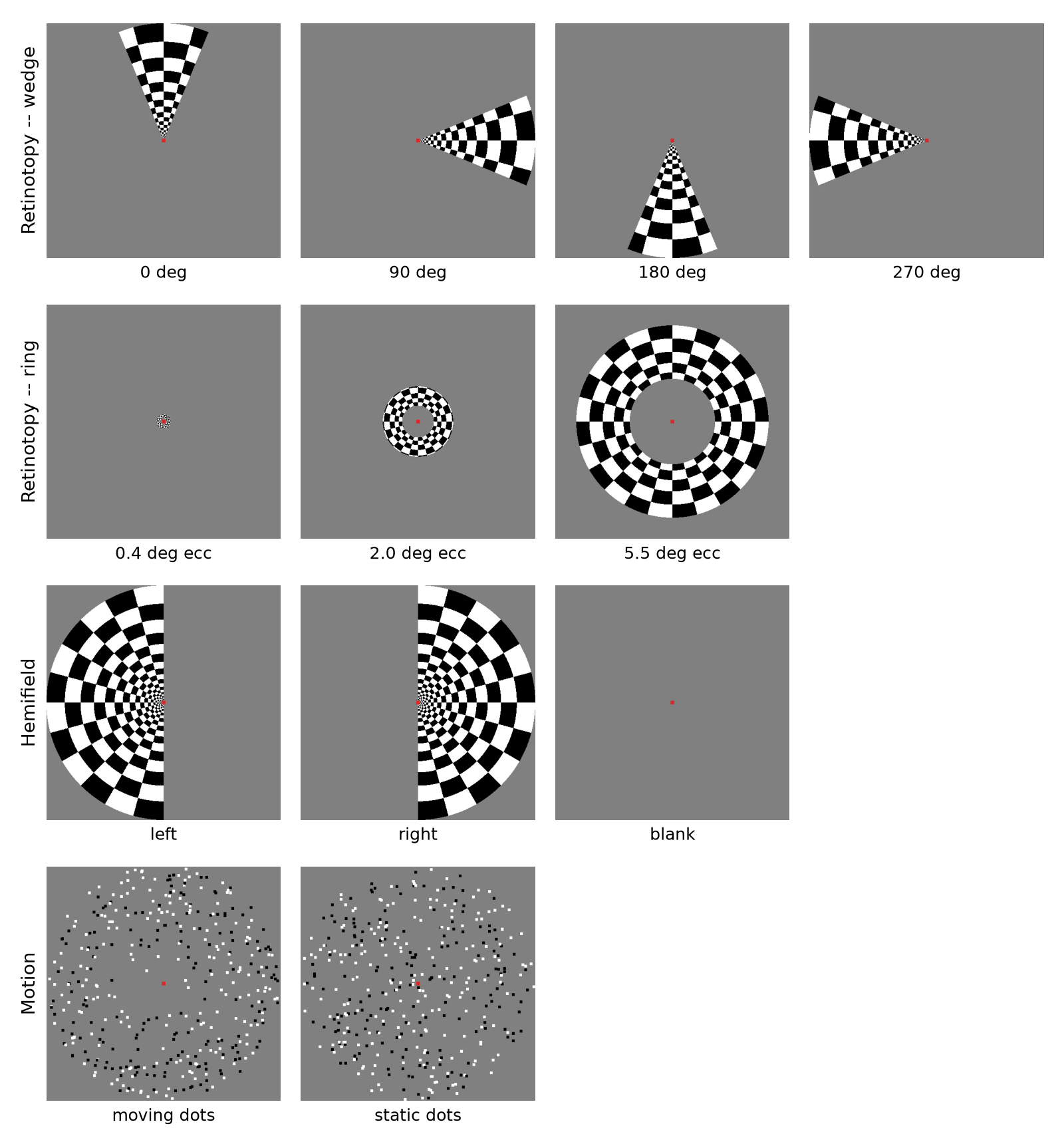}
\caption{Representative frames from each stimulus video. \textbf{Retinotopy -- wedge:} a
45\textdegree\ checkerboard wedge rotating through a full cycle (shown at 0\textdegree,
90\textdegree, 180\textdegree\ and 270\textdegree). \textbf{Retinotopy -- ring:} a
checkerboard annulus expanding logarithmically from near fixation (0.4\textdegree) to the
screen edge (shown at 0.4\textdegree, 2.0\textdegree\ and 5.5\textdegree\ eccentricity).
\textbf{Hemifield:} left, right and blank blocks of a flickering checkerboard restricted to
one visual hemifield. \textbf{Motion:} radially moving vs.\ static random-dot fields used as
an MT+ localizer. All frames are shown at native 512$\times$512 resolution with the central
fixation marker.}
\label{fig:stimuli}
\end{figure}

I generated the classic stimuli used to map each system experimentally: counter-rotating
checkerboard wedges and expanding/contracting rings (retinotopy), a left/right hemifield
flicker contrast, radially moving vs.\ static random-dot fields (motion), and ascending and
descending logarithmic tone sweeps spanning 175~Hz--5.6~kHz in 10 six-semitone bands
(tonotopy). Full stimulus parameters are in \texttt{config.yaml} of the project
repository.\footnote{\url{https://github.com/jakubstrawa1/audit-multimodal-brain-response}}
Figure~\ref{fig:stimuli} shows representative frames extracted from each stimulus video.

\subsection{Model and inference}
I ran the released \texttt{facebook/tribev2} checkpoint (video encoder: V-JEPA~2 giant;
audio encoder: Wav2Vec-BERT~2.0) on each stimulus, bypassing TRIBE's text/transcription path
entirely since no stimulus contains speech. Predictions were produced per-vertex at 1~Hz
(TR~=~1~s), with TRIBE's built-in 5-second hemodynamic-lag correction already applied.

\subsection{Fitting predicted responses}
I fit the same models to TRIBE's \emph{predicted} time courses that are conventionally fit
to real fMRI time courses:
\begin{itemize}
  \item \textbf{Retinotopy:} a 2D Gaussian population receptive field (pRF; position and
    size) via grid search over wedge and ring responses jointly \citep{dumoulin2008population},
    plus Fourier power at the wedge frequency ($f$) vs.\ twice that frequency ($2f$) to
    separate position-coding from orientation-coding responses.
  \item \textbf{Hemifield:} contralateral $-$ ipsilateral response effect size.
  \item \textbf{Motion:} moving $-$ static response effect size.
  \item \textbf{Tonotopy:} Gaussian tuning curves on log frequency (best frequency, tuning
    width) fit jointly to ascending and descending sweeps.
\end{itemize}
An additional $-3$ to $+8$~s lag search was allowed on top of TRIBE's built-in shift, since
the model's predicted auditory responses lag the stimulus by a further 2--3~s (Section
\ref{sec:tonotopy-result}).

\subsection{Reference maps and null models}
Fitted maps were compared against published human atlases on the same fsaverage5 mesh: the
Benson retinotopy template for polar angle, eccentricity and V1--V3 boundaries
\citep{benson2014correction,benson2018bayesian}, the Glasser HCP-MMP parcellation for MT and
MST \citep{glasser2016multimodal}, and the posteromedial-to-anterolateral axis of A1 (the
best-replicated human tonotopic gradient) for tonotopy \citep{saenz2014tonotopic,
moerel2014anatomical}. Because TRIBE is deterministic, there is no trial-to-trial sampling
noise to test against; instead I ask whether a map's spatial layout is aligned with known
anatomy, using a spin test that rotates each map on the sphere while preserving its spatial
autocorrelation \citep{alexanderbloch2018testing}, and, for tonotopy, a stimulus-label
permutation that shuffles which frequency band is which.

\subsection{Positive/negative control}
Before running the real model, I validated the entire stimulus-generation, fitting and
evaluation pipeline against a stand-in model with two configurable response modes: a
\emph{retinotopic+tonotopic} mode that plants known pRFs (from the Benson template) and an
A1 frequency gradient, and a \emph{pooled+flat} mode where visual vertices see only global
image statistics (stimulated area, wedge orientation) and auditory vertices see only
sound on/off -- i.e., exactly what TRIBE's architecture is expected to produce if H1 and H4
(null) hold. This lets me confirm the pipeline recovers a map when one is planted, and
correctly reports its absence when it is not.

\section{Results}

\subsection{Pipeline validation (stand-in model)}
\label{sec:standin-result}

Table~\ref{tab:standin} confirms the analysis pipeline has the sensitivity and specificity
needed to detect each type of map. The retinotopic+tonotopic stand-in recovers strong,
significant contralaterality, polar-angle, and frequency-gradient effects; the pooled+flat
stand-in -- built to produce exactly the responses TRIBE's architecture is expected to
produce under H1/H4 -- shows none of these effects (R1, R2, T1 near zero, R2b at chance),
with the single exception of a weak, inconsistent eccentricity correlation (R3; significant
in one of four random seeds), which is why eccentricity claims require the area-matched ring
control (Section~\ref{sec:controls}).

\begin{table}[h]
\centering
\begin{tabular}{>{\raggedright\arraybackslash}p{0.34\linewidth} >{\centering\arraybackslash}p{0.30\linewidth} >{\centering\arraybackslash}p{0.28\linewidth}}
\toprule
\textbf{Test} & \textbf{Retinotopic+tonotopic stand-in} & \textbf{Pooled+flat stand-in} \\
\midrule
R1 V1 contralateral effect ($d$)            & 2.73 (spin $p=0.007$)  & 0.01 ($p=0.12$) \\
R2 polar angle vs.\ template (circular $r$)  & 0.997 ($p=0.001$)       & $-$0.01 ($p=0.55$) \\
R2b pRFs on contralateral side               & 100\%                   & 52--57\% (chance) \\
R3 eccentricity vs.\ template (Spearman)      & 0.991 ($p=0.001$)       & 0.01--0.12$^{*}$ \\
R4 wedge power at $f$ vs.\ $2f$                & 0.60 (position-driven)  & 0.48 (orientation-driven) \\
M1 MT+ moving $-$ static ($d$)               & 3.95 ($p=0.001$)        & 3.90 (planted in both) \\
T1 A1 frequency gradient (Spearman)          & $-$0.99 (spin $p=0.006$; label $p=0.007$) & $+$0.12 ($p\approx 0.75$) \\
T2 best-frequency spread (octaves)           & 2.0                     & 0.0 \\
\bottomrule
\end{tabular}
\caption{Stand-in model validation. $^{*}$Across four random seeds of the pooled model, the
eccentricity correlation (R3) was significant in only one (0.12, $p=0.04$); see
Section~\ref{sec:controls}.}
\label{tab:standin}
\end{table}

\begin{figure}[H]
\centering
\includegraphics[width=0.48\linewidth]{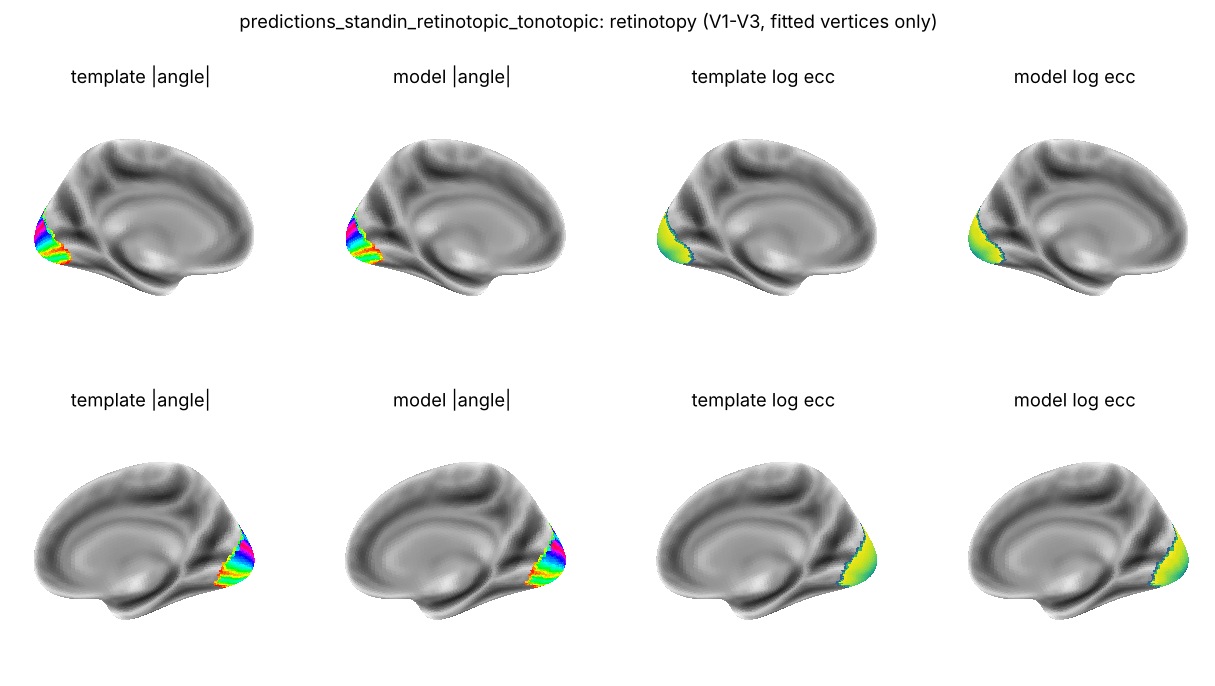}
\includegraphics[width=0.48\linewidth]{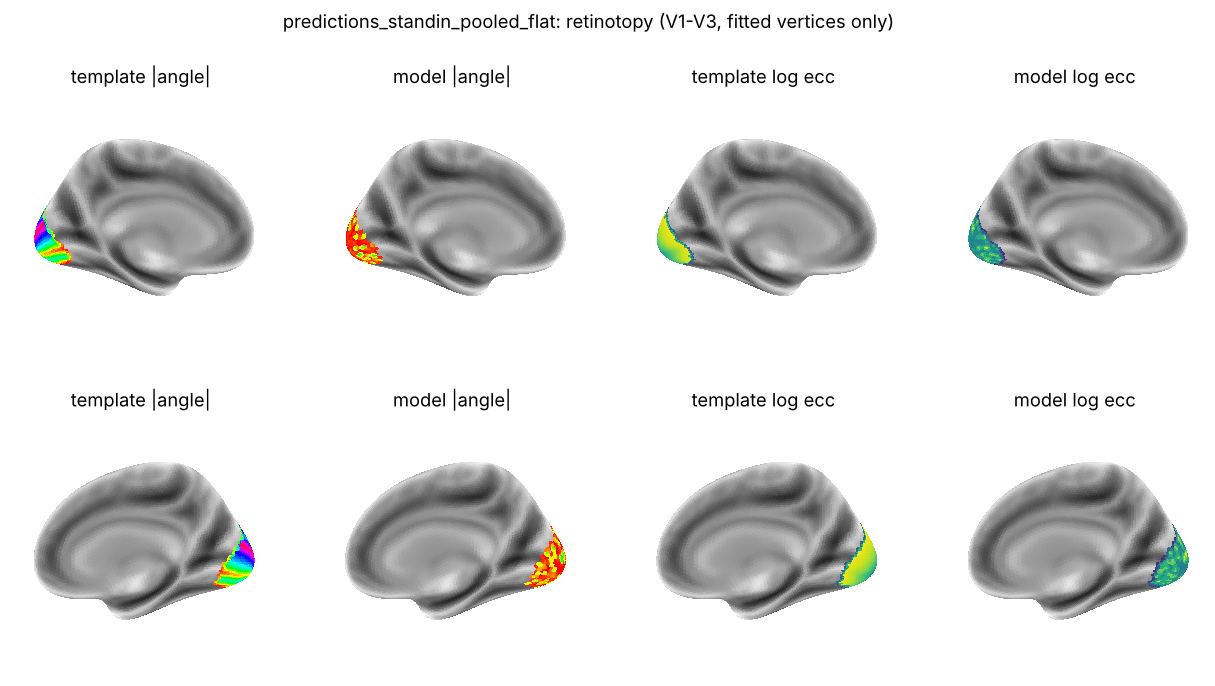}
\caption{Fitted pRF maps from the stand-in model. \emph{Left:} retinotopic+tonotopic mode
(positive control) recovers the planted polar-angle/eccentricity structure. \emph{Right:}
pooled+flat mode (negative control, matching TRIBE's expected architecture-level behavior)
shows no positional structure.}
\label{fig:standin}
\end{figure}

\subsection{Tonotopy: TRIBE v2 predicts no tonotopic gradient}
\label{sec:tonotopy-result}

\begin{figure}[H]
\centering
\includegraphics[width=0.75\linewidth]{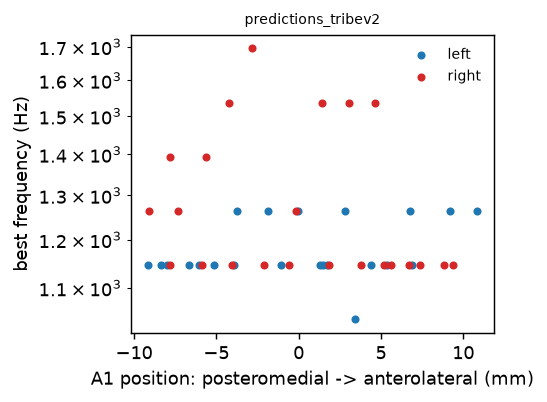}
\caption{Best-frequency map and gradient along A1 from real TRIBE v2 predictions on the
tone-sweep stimuli. No posteromedial-to-anterolateral gradient is visible.}
\label{fig:tonotopy}
\end{figure}

I ran the real TRIBE v2 model on the ascending and descending tone-sweep stimuli. Auditory
cortex responds robustly to the tones: every core and belt vertex is well fit (T0: 100\%
clear the fit-quality threshold), and A1 is strongly suppressed during silence
($z \approx -2.1$). However, I find no evidence of a tonotopic map: the spread of best
frequencies across core and belt is only 0.141~octaves (T2), against 2.0~octaves in the
planted-map stand-in control, and there is no gradient along A1 (T1: Spearman
$\rho = -0.067$; spin $p=0.418$; label-permutation $p=0.343$). The posteromedial and
anterolateral halves of A1 show near-identical band-by-band response profiles
(Figure~\ref{fig:tonotopy}). This result was first observed in a smaller pilot run on the
audio stimuli alone (T2 $=0.14$ octaves, $\rho=-0.07$) and is now confirmed on the full,
six-stimulus run reported here -- the numbers are essentially unchanged. Predicted responses
peak 2--3~s later than TRIBE's built-in 5-second hemodynamic-lag correction assumes, which I
accommodate with the widened lag search described in Methods.

\textbf{H4 verdict: no tonotopic gradient.} H4 was stated as an open question; the answer
is that TRIBE v2 predicts none. Because my main tones are equal in amplitude rather than in
perceived loudness, the uniform best-frequency preference could partly reflect loudness. The
headline finding does not depend on this, since every vertex is exposed to identical sounds,
and the loudness-equalized control (Section~\ref{sec:controls}) confirms that the gradient is
still absent under equal loudness. I do not interpret the specific preferred frequency
($\sim$1.2~kHz) further.

\subsection{Retinotopy: no positional code survives}
\label{sec:retinotopy-result}

\begin{figure}[H]
\centering
\includegraphics[width=0.9\linewidth]{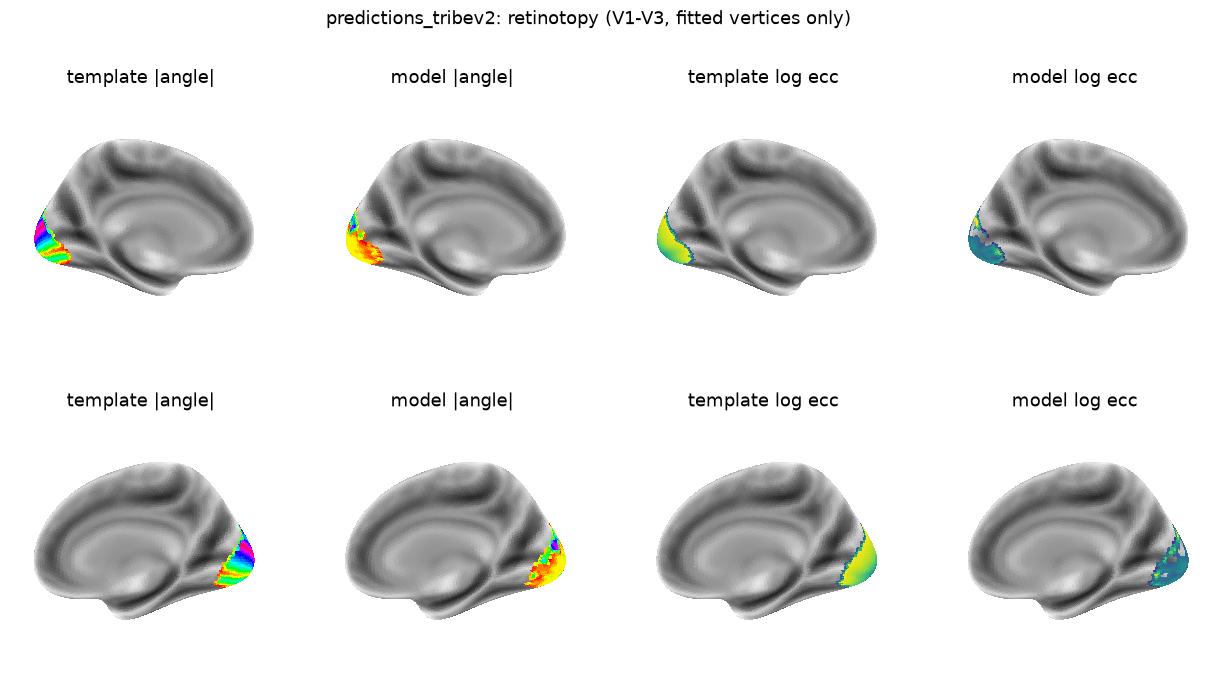}
\caption{Fitted pRF maps (|polar angle|, log eccentricity) from real TRIBE v2 predictions on
the wedge and ring stimuli (two hemispheres shown), compared against the Benson template.
The model's maps (right-hand columns within each pair) show no structure resembling the
template's smooth angular/eccentricity gradients.}
\label{fig:retinotopy}
\end{figure}

Table~\ref{tab:realresults} reports the full battery on real TRIBE v2 predictions for all
four video stimuli. Fitted pRFs are well-behaved in the descriptive sense (R0: 100\% of
V1--V3 vertices clear the fit-quality threshold), but show no positional structure: the
contralateral hemifield effect is near zero and non-significant (R1 $d=0.048$, $p=0.237$),
polar angle does not correlate with the Benson template (R2 circular $r=-0.024$, $p=0.595$),
and eccentricity is, if anything, weakly negatively correlated with the template rather than
positively (R3 Spearman $=-0.129$, $p=0.724$) \emph{at this default ring geometry} -- but this
specific null is not stable: the area-matched and stimulus-size controls
(Section~\ref{sec:controls}) flip R3 to significant and positive ($r=0.44$--$0.76$,
$p\le0.014$) whenever the ring's rendering geometry changes, while R1/R2/R2b stay robustly
null throughout. I treat the eccentricity metric (R3) as unresolved rather than a clean null;
see Section~\ref{sec:controls} and \ref{sec:discussion}. R2b (fraction of pRFs landing in the
geometrically correct contralateral hemifield) is 0.61, closer to chance (0.5) than to the
stand-in positive control's 1.0. Figure~\ref{fig:retinotopy} makes this visible directly: the
model's fitted polar-angle and eccentricity maps are nearly uniform in color across V1--V3,
in clear contrast to the smoothly varying template maps beside them.

One result complicates a fully clean null, however: R4 (the share of wedge-locked power at
the stimulus frequency $f$ versus its first harmonic $2f$) is 0.744 -- above the 0.5 midpoint
and nominally in the ``position-coding'' direction, not the ``orientation-coding'' direction the
interpretation key (Table~\ref{tab:interpretation}) associates with a clean H1 confirmation.
Taken together with R1--R3, I read this as: TRIBE v2's predicted responses are phase-locked
to the wedge's periodic motion, but not in a way that decodes to a coherent, template-aligned
polar-angle map -- i.e., periodicity survives spatial averaging even though position does
not. I discuss this discrepancy further in Section~\ref{sec:discussion}.

\textbf{H1 verdict: confirmed} for the position code H1 actually concerns -- contralaterality
and polar angle. No contralateral hemifield effect and no polar-angle map, robustly so across
every control in Section~\ref{sec:controls} (ring geometry, wedge width, audio). This matches
the architecture-level prediction that spatially-averaged video features cannot support a
retinotopic code. Two caveats are carried forward rather than folded into the headline claim:
R4 (above) and, more importantly, R3/eccentricity, which is \emph{not} a stable null -- it
flips to significant and template-aligned whenever the ring's rendering geometry changes
(Section~\ref{sec:controls}), consistent with the H2 prediction of an
eccentricity-\emph{like} signal driven by global image statistics rather than a true positional
code. I treat eccentricity as an open question, not part of the confirmed H1 null.

\textbf{H2 verdict: consistent with the prediction}, in the ``partly present'' form stated in
Table~\ref{tab:hypotheses}. There is no stable, geometry-invariant eccentricity code (R3 is
null at the default ring geometry, and R1/R2/R2b show no positional code of any other kind),
but an eccentricity-\emph{like} signal appears whenever the ring geometry changes
(Section~\ref{sec:controls}), most plausibly carried by global image statistics that covary
with that geometry.

\subsection{Motion: TRIBE v2 correctly localizes motion selectivity to MT+}
\label{sec:motion-result}

In clear contrast to retinotopy, the motion contrast is strong and precisely localized. The
moving-minus-static effect size is large and significant both within MT+ (M1 $d=4.31$,
spin $p=0.001$) and, more weakly, across the rest of cortex (M1b $d=0.67$), but the top
response is sharply concentrated: 22.0\% of the single most motion-selective 1\% of all
cortical vertices fall inside the (anatomically tiny) MT+ parcel, a more than 50-fold
enrichment over MT+'s 0.42\% share of cortex (M2). Figure~\ref{fig:motion} shows the
moving-static contrast map directly: a well-defined cluster in the expected posterolateral
location on each hemisphere.

\textbf{H3 verdict: confirmed.} TRIBE v2 represents motion information, and the response is
anatomically localized to the correct region, not merely present somewhere in cortex.

\subsection{Hemifield contrast map}

\begin{figure}[H]
\centering
\includegraphics[width=0.9\linewidth]{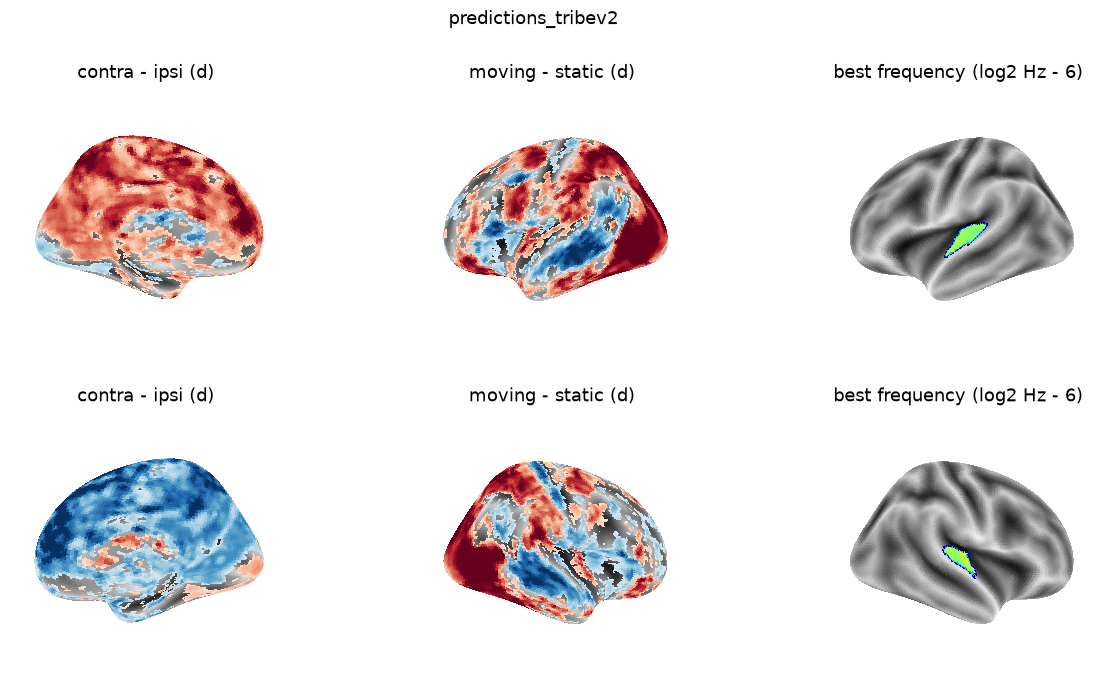}
\caption{Left: contralateral-minus-ipsilateral hemifield contrast (two hemispheres). Middle:
moving-minus-static motion contrast, showing a focal cluster in the expected MT+ location.
Right: best-frequency map in and around A1, showing no spatial gradient (Section
\ref{sec:tonotopy-result}).}
\label{fig:motion}
\end{figure}

Figure~\ref{fig:motion} (left column) also shows the raw contralateral-minus-ipsilateral
hemifield contrast map prior to V1--V3 masking; the diffuse, non-focal pattern across much of
lateral and medial cortex is consistent with the null R1 statistic reported above and with
the absence of a clean retinotopic code in early visual cortex specifically.

\begin{table}[h]
\centering
\small
\begin{tabular}{lcc}
\toprule
\textbf{Test} & \textbf{Value} & \textbf{$p$ (vs.\ spin/label null)} \\
\midrule
R0 V1--V3 vertices with pRF fit $r \geq$ threshold (fraction) & 1.00 & -- \\
R1 V1 contralateral effect ($d$) & 0.048 & 0.237 \\
R2 polar angle vs.\ template (circular $r$) & $-$0.024 & 0.595 \\
R2b pRFs in correct contralateral hemifield (fraction) & 0.612 & -- (0.5 = chance) \\
R3 eccentricity vs.\ template (Spearman) & $-$0.129 & 0.724 \\
R4 wedge power at $f$ / ($f$+$2f$), V1--V3 median & 0.744 & -- ($>$0.5: position-coding) \\
M1 moving $-$ static effect in MT+ ($d$) & 4.312 & 0.001 \\
M1b moving $-$ static effect, rest of cortex ($d$) & 0.672 & -- \\
M2 top-1\% motion vertices inside MT+ (fraction) & 0.220 & -- (vs.\ 0.0042 chance share) \\
T0 auditory vertices with tuning fit $r \geq$ threshold (fraction) & 1.00 & -- \\
T1 A1 best frequency vs.\ Heschl's axis (Spearman) & $-$0.067 & 0.418 (spin) / 0.343 (label) \\
T2 spread of best frequency, core+belt (IQR, octaves) & 0.141 & -- \\
\bottomrule
\end{tabular}
\caption{Full results on real TRIBE v2 predictions, all six stimuli (four video, two audio).}
\label{tab:realresults}
\end{table}

\begin{table}[h]
\centering
\begin{tabular}{p{0.45\linewidth}p{0.48\linewidth}}
\toprule
\textbf{Pattern} & \textbf{Meaning} \\
\midrule
R1 $\approx$ 0, R2 $\approx$ 0, R2b $\approx$ 0.5, R4 $<$ 0.5 &
  H1 confirmed: no positional code; responses follow orientation and stimulated area, not
  location. In-silico experiments depending on visual-field position are invalid with
  TRIBE v2. \\
R1 $>$ 0 and R2b $>$ 0.5, but R2 weak &
  Coarse left/right information survives averaging; fine polar angle does not. \\
R3 $>$ 0 while R1/R2 $\approx$ 0 &
  Eccentricity from global image statistics (ring size), not true position; confirm with
  the area-matched ring control. \\
M1 large, well above MT+'s share of cortex &
  H3 confirmed: motion selectivity localizes correctly. \\
T1 $<$ 0 with both $p$-values small, T2 $>$ 1~octave &
  H4 confirmed: a genuine tonotopic gradient, learned without explicit tone training. \\
T2 $\approx$ 0 &
  Auditory vertices share one tuning: no tonotopy (as observed; Section
  \ref{sec:tonotopy-result}). \\
\bottomrule
\end{tabular}
\caption{Interpretation key applied to the full result set. Note: the real TRIBE v2 data
match the ``H1 confirmed'' row on R1/R2/R2b but \emph{not} on R4 ($0.744 > 0.5$); R4 is
treated as a separate, unresolved caveat rather than a condition on the H1 verdict (see
Section~\ref{sec:retinotopy-result}).}
\label{tab:interpretation}
\end{table}

\section{Controls and robustness}
\label{sec:controls}

\begin{enumerate}
  \item \textbf{Silent audio vs.\ no audio.} My video stimuli carry a silent audio track by
    default; I re-ran video experiments with the audio input dropped entirely. R1 and R2
    remain non-significant with audio dropped (R1 $d=0.045$, $p=0.279$; R2 $r=-0.031$,
    $p=0.540$; vs.\ $d=0.048$, $p=0.237$ and $r=-0.024$, $p=0.595$ with audio present), so the
    retinotopic null is unaffected. The MT+ motion effect also survives (M1 $d=3.25$,
    $p=0.001$ vs.\ $d=4.31$, $p=0.001$ with audio), though both its effect size and its
    top-1\%-vertex enrichment in MT+ roughly halve without the silent audio channel (M2
    10.7\% vs.\ 22.0\%) -- a real quantitative difference, but not one that changes which
    hypotheses are confirmed or disconfirmed.
  \item \textbf{Eccentricity vs.\ area.} A ring control varying position while holding
    stimulated area constant tests whether the R3 result (Section~\ref{sec:retinotopy-result})
    reflects position or area. Holding area constant does \emph{not} null the effect: R3
    becomes significant and positive (Spearman $r=0.469$, $p=0.014$), in contrast to the
    non-significant, slightly negative baseline ($r=-0.129$, $p=0.724$). This rules out
    stimulated area alone as the explanation and is discussed together with the stimulus-size
    controls below.
  \item \textbf{Stimulus size.} Repeating the analysis with \texttt{max\_ecc\_deg} of 5 and
    20 (vs.\ the default 10) was expected to leave rank-based results unchanged. This holds for
    R1 and R2, which stay non-significant at both scales (R1 $d=0.047$/$0.044$, $p=0.231$/$0.240$; R2
    $r=0.028$/$-0.0004$, $p=0.497$/$0.478$), but R3 becomes significant and positive at both
    --- $r=0.440$, $p=0.012$ at 5\textdegree\ and $r=0.755$, $p=0.001$ at 20\textdegree\ ---
    the same qualitative flip seen in the area control above. Across all three controls that
    regenerate the ring stimulus with altered geometry (area-matched, 5\textdegree,
    20\textdegree), R3 flips from a non-significant, slightly negative baseline to a
    significant, positive, template-aligned correlation, while R1/R2/R2b stay robustly null
    throughout. I read this as consistent with the H2 prediction (an eccentricity-\emph{like}
    gradient driven by global image statistics that covary with ring geometry) rather than a
    genuine, geometry-invariant positional code. Since the area-matched control already rules
    out stimulated area alone, other geometric properties (for example ring width, check size
    or proximity to the frame edge) are candidates; these controls do not identify which one is
    responsible -- see the discussion of R3 in Section~\ref{sec:retinotopy-result} and
    \ref{sec:discussion}.
  \item \textbf{Wedge width and check size.} Repeating the wedge experiment at 30\textdegree\
    and 90\textdegree\ (vs.\ the default 45\textdegree) leaves H1's central null intact: polar
    angle does not correlate with the template at either width (R2 $r=-0.026$, $p=0.569$ at
    30\textdegree; $r=0.013$, $p=0.557$ at 90\textdegree; vs.\ $r=-0.024$, $p=0.595$ at
    45\textdegree), and the R4 position-vs-orientation caveat (Section~\ref{sec:retinotopy-result})
    persists essentially unchanged at all three widths (R4 $=0.734$, $0.759$, $0.744$
    respectively).
  \item \textbf{Loudness-equalized tones.} An ISO~226 equal-loudness version of the tone
    sweeps tests whether the missing gradient could be a loudness artifact (``best frequency
    = loudest frequency''; Section~\ref{sec:tonotopy-result}). The null persists under equal
    loudness: best frequency still does not track Heschl's axis
    (Spearman $r=-0.081$, spin $p=0.405$, label $p=0.340$; vs.\ $r=-0.067$, spin $p=0.418$,
    label $p=0.343$ with equal-amplitude tones), confirming the absent tonotopic gradient is
    not a loudness artifact.
\end{enumerate}

\section{Discussion}
\label{sec:discussion}

Across all four hypotheses, TRIBE v2 shows a consistent pattern: codes that depend on
\emph{where} something is (visual-field position, tonotopic frequency position) are absent,
while a code that depends on \emph{what kind of feature} is present without requiring
position (motion) survives and is correctly localized. H1 (retinotopy) is confirmed: as the
architecture predicts, the spatially averaged video stream carries no usable positional code.
H4 (tonotopy), stated as an open question, resolves as absent -- more surprisingly, since the
audio pathway has no equivalent spatial-pooling step. The tonotopic null therefore cannot be
explained by the video pathway's pooling. It may reflect what the model's audio training data
(mostly speech and film sound) taught it to represent, the audio encoder's features, or how
the response head aggregates sequence-level audio features; this audit does not distinguish
between these. H3 (motion) is confirmed cleanly: MT+ selectivity
is both strong and anatomically precise (M2), showing that TRIBE v2 can support
in-silico experiments that depend on \emph{feature identity} even in domains where
position-dependent experiments fail.

The one result that resists a clean story is R4 (Section~\ref{sec:retinotopy-result}):
wedge-locked periodic power is present (0.744, above the position-coding threshold) despite
the complete absence of a coherent polar-angle map (R2) or contralaterality effect (R1). I
see two readings. First, R4 may simply be a weaker, noisier signal than R1--R3: it requires
only that \emph{some} wedge-synchronized response exists somewhere in V1--V3, not that it be
spatially organized, so it is a much lower bar to clear and may reflect overall
stimulus-locked arousal/attention-like modulation rather than a retinotopic code. Second, it
leaves open the possibility that some coarse spatial information survives averaging in a form
the circular-correlation metric (R2) is not sensitive to -- for instance, information about
stimulated area or aggregate screen-side energy that correlates with wedge phase without
reconstructing a true angular map. Distinguishing these two readings is a natural next step
and does not change the paper's central claim, since neither reading implies a usable
positional (polar-angle or eccentricity) code.

R3 (eccentricity) shows a related pattern and is, if anything, a cleaner case of the same
phenomenon: it is a non-significant, slightly negative null at the default ring geometry, but
becomes a significant, template-aligned correlation under every control that regenerates the
ring with different geometry (area-matched, 5\textdegree, 20\textdegree; Section
\ref{sec:controls}). Because its sign and significance depend on how the ring is rendered,
and holding stimulated area constant does not remove it, R3 does not behave like a stable
positional code. I read it as consistent with the H2 prediction -- an eccentricity-\emph{like}
gradient carried by global image statistics that covary with ring geometry -- rather than
evidence that TRIBE v2 retains true positional information, although these controls do not
identify the statistic responsible, and a weak genuine eccentricity signal cannot be fully
excluded. This matters for practitioners: an apparent eccentricity gradient in TRIBE v2's
predictions is not by itself evidence of a retinotopic code, and should be checked across
several stimulus geometries before being interpreted.

For practitioners using TRIBE v2 for in-silico neuroscience, the practical implication is
direct: experiments that depend on visual-field or tonotopic position (e.g., contralateral
hemifield localizers, pRF mapping, frequency-selective auditory designs) should not be
trusted with this model, while experiments that depend on category- or feature-level
selectivity (as in the original paper's face/place/body/word localizers, and as I
demonstrate for motion) remain viable. For vision, the missing positional code is most
likely architectural: the video features are averaged over space before the response head,
so further training of the current design would not be expected to restore it, and a TRIBE
version aiming to support position-based in-silico experiments would likely need to preserve
spatial structure through to the response head. For audition the cause remains open, as
discussed above. Ablations that modify the pooling step or the audio features are the
natural way to test these explanations directly.

\section{Limitations}

\begin{itemize}
  \item \textbf{Resolution.} fsaverage5 vertices are spaced approximately 4~mm apart; A1
    comprises only about 45 vertices in total (22 left, 23 right) at this resolution,
    limiting sensitivity to fine-grained tonotopic structure.
  \item \textbf{Average subject.} TRIBE predicts a group-average subject's response, so
    fitted maps are necessarily compared against group-level templates rather than
    individual-subject maps.
  \item \textbf{Visual angle is undefined for TRIBE.} I define the stimulus frame to span
    \texttt{max\_ecc\_deg} degrees (10 by default). The contralaterality and polar-angle tests
    (R1, R2) are unaffected by this choice, but the eccentricity correlation (R3) is not
    (Section~\ref{sec:controls}), which is one reason R3 is treated as unresolved.
  \item \textbf{Mechanisms are inferred, not tested.} The architectural explanation for the
    visual null comes from the published model description; I did not modify the model to
    test it.
  \item \textbf{Tonotopic axis.} The orientation of tonotopic gradients relative to
    Heschl's gyrus remains debated in the human-neuroimaging literature; T1 specifically
    tests the posteromedial-high/anterolateral-low pattern, while T2 (best-frequency spread)
    is direction-free.
\end{itemize}

\section*{Acknowledgments}
This work used approximately 8 hours of H100 GPU time, including the control analyses,
self-funded by the author with no external funding or grant support.

\bibliographystyle{apalike}
\bibliography{refs}

\end{document}